\documentclass[%
superscriptaddress,
reprint,
amsmath,
amssymb,
aps,
pra,
floatfix
]{revtex4-2}

\usepackage{xcolor}
\usepackage{graphicx}
\usepackage{bm}
\usepackage{array}

\begin{document}

\title{
Long nuclear spin coherence times for 
molecules trapped in high-purity solid parahydrogen
}

\author{Alexandar P. Rollings}
\affiliation{Department of Physics, University of Nevada, Reno NV 89557, USA}
\author{Jonathan D. Weinstein}
\email{weinstein@physics.unr.edu}
\homepage{https://www.weinsteinlab.org}
\affiliation{Department of Physics, University of Nevada, Reno NV 89557, USA}

\begin{abstract}
We measure the ensemble transverse relaxation time ($T_2^*$) and spin-echo coherence time ($T_2$) of the proton spin of HD molecules  trapped in solid parahydrogen. By using high-purity parahydrogen matrices, we are able to measure significantly longer $T_2$ and $T_2^*$ times than seen in prior work.  
We also measure the longitudinal spin relaxation time $T_1$. We examine how these parameters scale with the matrix purity and find limits on the coherence time from the parahydrogen matrix itself.
\end{abstract}

\maketitle

\section{Introduction}

Molecules have demonstrated tremendous power as sensors for physics beyond the Standard Model of particle physics \cite{safronova2018search}. 
Experiments with paramagnetic molecules have set the primary limits on the electron EDM \cite{acme2018improved, 13268}, providing constraints on symmetry-violating new physics  and 
on parameters of both standard-model physics and beyond-standard-model physics \cite{chupp2019electric}. %
Similarly, experiments using diamagnetic molecules with nuclear spin can be used to probe symmetry-violating new physics from the baryonic sector \cite{wilkening1984search, cho1991search, grasdijk2021centrex, yu2021probing, fan2021optical, chen2024relativistic, engel2025nuclear}. 

As has been previously noted \cite{PhysRevLett.97.063001, vutha2018oriented, PhysRevA.98.032513}, trapping molecules in a matrix may offer significant improvements in the statistical sensitivity of such experiments, due to the large numbers of molecules that can be trapped. 
For experiments using diamagnetic molecules to search for symmetry-violating new physics from the nucleus, the nuclear spin ensemble transverse relaxation time $T_2^*$ is a critical figure of merit, with long $T_2^*$ times offering better sensitivity \cite{engel2025nuclear, safronova2018search}.

Typically, without techniques such as magic-angle-spinning \cite{Andrew1981MagicAngleSpinning}, $T_2^*$ is extremely short for molecules in solids. This is primarily due to broadening from
inhomogenous molecular orientations and conformations, as well as magnetic interactions with other magnetic dipoles in the matrix  \cite{kittel1953dipolar}. By trapping diatomic molecules in solid parahydrogen, it may be possible to eliminate these dominant broadening mechanisms. First, IR spectroscopy of diatomic molecules in solid parahydrogen indicates they can freely rotate \cite{Momose1998}, eliminating broadening from inhomogenous molecular orientations. 
Second, diatomic molecules only have a single conformation. Finally, parahydrogen --- the $I=0$ state of  H$_2$ ---  is magnetically pure. Thus magnetic dipolar interactions should go to zero in the limit of low densities of the dopant molecules and low densities of the $I=1$ orthohydrogen state of H$_2$.

Prior work in solid parahydrogen measured the NMR properties of the proton of the  HD molecule down to an orthohydrogen fraction $X=2 \times 10^{-3}$ \cite{delrieu1981quantum, washburn1981nmr,  zhou1990nmr, kisvarsanyi1992nuclear}.
Here, we extend that work to $X=10^{-6}$.

\section{Apparatus}

\begin{figure}[!ht]
			\includegraphics[width=\linewidth]{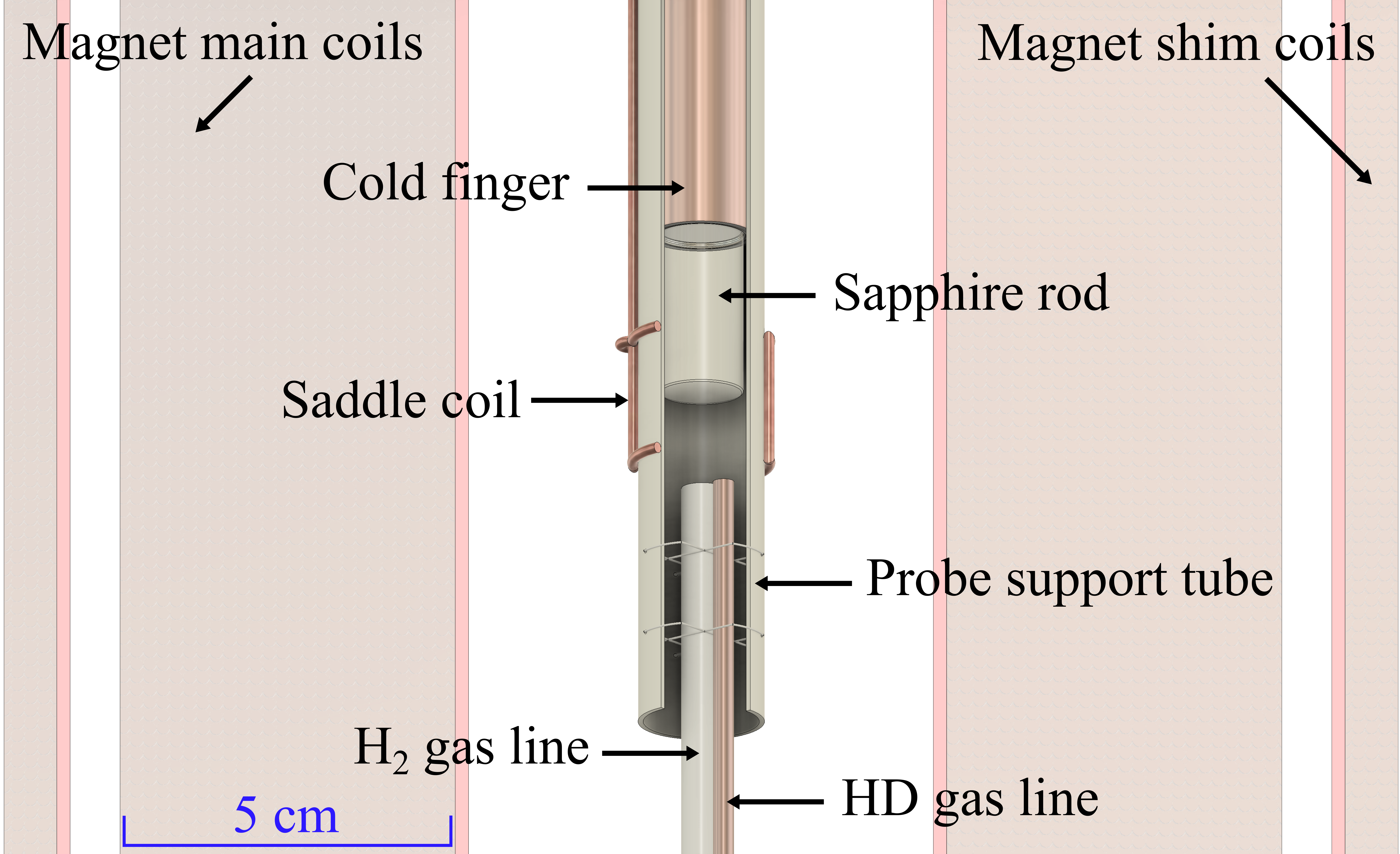}
				\caption[]{ \label{fig:apparatus}
                Scale-accurate schematic of the apparatus, as described in the text.
                The saddle coil and support tube are shown in cutaway views, and the cylindrically-symmetric magnet is shown in cross-section.
				}
\end{figure}

A schematic of the experimental apparatus is shown in Fig. \ref{fig:apparatus}. A detailed description can be found in references \cite{RSI_NMR_Apparatus} and  \cite{RollingsDissertation}. In brief, the apparatus consists of a single vacuum chamber and a single closed-cycle pulse-tube refrigerator that provides cooling for all the cryogenic components.

The samples are grown on a sapphire rod attached to the end of a copper cold finger. Sapphire was chosen for its high thermal conductivity and low magnetic susceptibility. 
Using literature values of the susceptibilities of OFE copper \cite{ekin2006} and high-purity sapphire \cite{guild_optics}, simulations indicate that the magnetic field $B$ in the sample should be uniform to within 340~ppb for a uniform applied field $H$ \cite{RSI_NMR_Apparatus, RollingsDissertation}. We note the field uniformity could be further improved with ``shimming'' of $H$.

The parahydrogen samples are grown on the sapphire by vapor deposition of natural-abundance hydrogen gas, with gas flow controlled by a mass flow controller. Purification of the parahydrogen gas is done by an ``in-line'' cryogenic catalyst as described in reference \cite{doi:10.1063/5.0049006}. By controlling the temperature of the catalyst, we can control the orthohydrogen fraction $X$ and achieve orthohydrogen fractions as low as $X = 1\times10^{-6}$ \cite{doi:10.1063/5.0049006}. 

The catalyst --- at sufficiently cold temperatures --- also performs isotopic separation of hydrogen and suppresses the HD component present in natural-abundance hydrogen gas \cite{doi:10.1063/5.0049006}. To both enable experiments at low $X$ and control the density of HD molecules in the matrix, we have a second gas line to dope our parahydrogen matrix with HD molecules. 

We also expect the catalyst to provide near-perfect chemical purification of the sample gas: only hydrogen and helium have sufficient vapor pressure to flow through the catalyst without freezing.

Samples are typically grown over a time period of 7 to 9 hours to a thickness of a few mm.
The copper cold finger temperature is 
4~K during sample deposition.
From prior work, we expect these growth conditions to produce polycrystalline samples  \cite{collins1996metastable}.
NMR measurements are also taken at 4~K. 

We operate the magnet in persistent current mode; typical magnetic bias fields are ${\sim}1.4$~T, giving a proton Larmor precession frequency of ${\sim}60$~MHz. 
Magnetic ``shim coils'' are used to adjust magnetic field gradients so as to provide a uniform magnetic  field. For the data presented here, we believe  $T_2^*$ is not limited by magnetic field inhomogeneities, as explained below.

The proton spins are driven with RF magnetic fields from a resonant ``saddle coil'' with a typical coupled $Q$ of 100 at base temperatures; this $Q$ is much smaller than the $Q$ of our nuclear spin ensemble. 
Typical Rabi frequencies are on the order of 3~kHz, which exceeds the proton linewidth in our sample.
The same saddle coil is used as a pickup coil to measure the Larmor precession of the spins. %

\section{Data}

\subsection{$T_2^*$}

\begin{figure}[!ht]
\includegraphics[width=\linewidth]{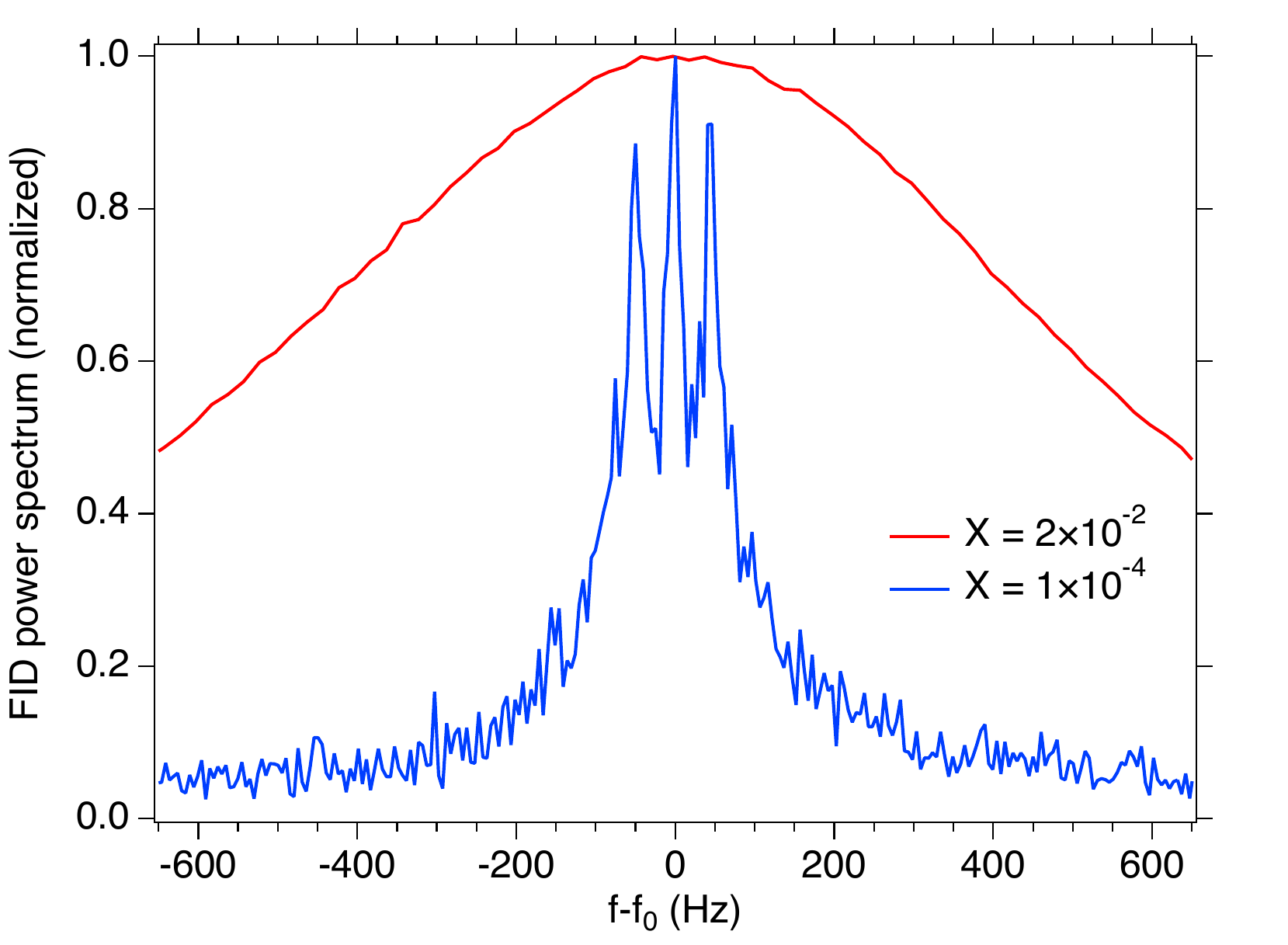}
	\caption[]{ \label{fig:FID}
		Power spectrum of the FID signal for two samples of differing orthohydrogen fraction $X$, as labeled. The broad and narrow lines have a frequency offset $f_0$ of 58.41 and 56.87 MHz respectively. 
				}
\end{figure}

We measure the ensemble transverse relaxation time $T_2^*$ by free induction decay (FID) measurements. We use a  $\pi/2$ pulse to induce precession of the H nuclei in our sample, and Fourier transform the resulting signal to get a power spectrum.
As seen in Fig. \ref{fig:FID}, the FID spectrum changes dramatically with the orthohydrogen fraction $X$. At high $X$ the line fits well to a single Lorentzian. At low $X$ the proton's triplet structure --- arising from electron-coupled intramolecular nuclear spin interactions with the deuteron \cite{ramsey1952interactions, ramsey1953electron} --- can be clearly resolved. 

These linewidths confirm that the HD molecule rotates within solid parahydrogen. If the molecular axis was fixed in space, intramolecular magnetic dipole--dipole interactions between the proton and deuteron in HD would give rise to splittings on the order of $10^5$~Hz. In the case of inhomogenous molecular orientations within the ensemble, this would result in broadening on the same order.

We determine $T_2^*$ from the linewidth of the FID spectrum. At high $X$ we fit the power spectrum to a single Lorentzian to determine the full width at half maximum (FWHM). In the case of low $X$ we fit the central region to the sum of three Lorentzians of equal FWHM. In both cases, we determine $T_2^*$ from $T_2^*=(\pi \cdot  \mathrm{FWHM})^{-1}$.  A summary of our $T_2^*$ measurements are presented in Fig. \ref{fig:T2sVsOrtho}, and are discussed in section \ref{sec:coherencetimes}.

We note that prior measurements of HD in solid hydrogen were unable to resolve the proton triplet \cite{reif1953nuclear, washburn1981nmr}. Here we measure a splitting of $J(H,D)=47.2 \pm 1.1$~Hz, differing significantly from the free molecule case of 43.1~Hz \cite{carr1952interaction, garbacz2014, garbacz2016indirect}. 
The difference is not entirely surprising: gas phase measurements are known to have a pressure dependence \cite{garbacz2014, garbacz2016indirect}, and liquid phase measurements similarly have a solvent-dependent shift \cite{benoit1967resonance, neronov1975nmr}. 

\subsection{$T_2$}

Because the ensemble transverse relaxation time $T_2^*$ can be limited by technical problems such as magnetic field inhomogeneities, we measure the spin-echo coherence time $T_2$ for comparison.
We induce precession with a $\pi/2$ pulse, followed by a refocusing pulse at a delay of $t/2$, and measure the resulting echo at time $t$. We fit the power spectrum amplitude to the functional form $A \cdot \exp(-2t/T_2) + y_0$ to extract the spin-echo $T_2$ \cite{T2footnote}.

\begin{figure}[!ht]
    \includegraphics[width=\linewidth]{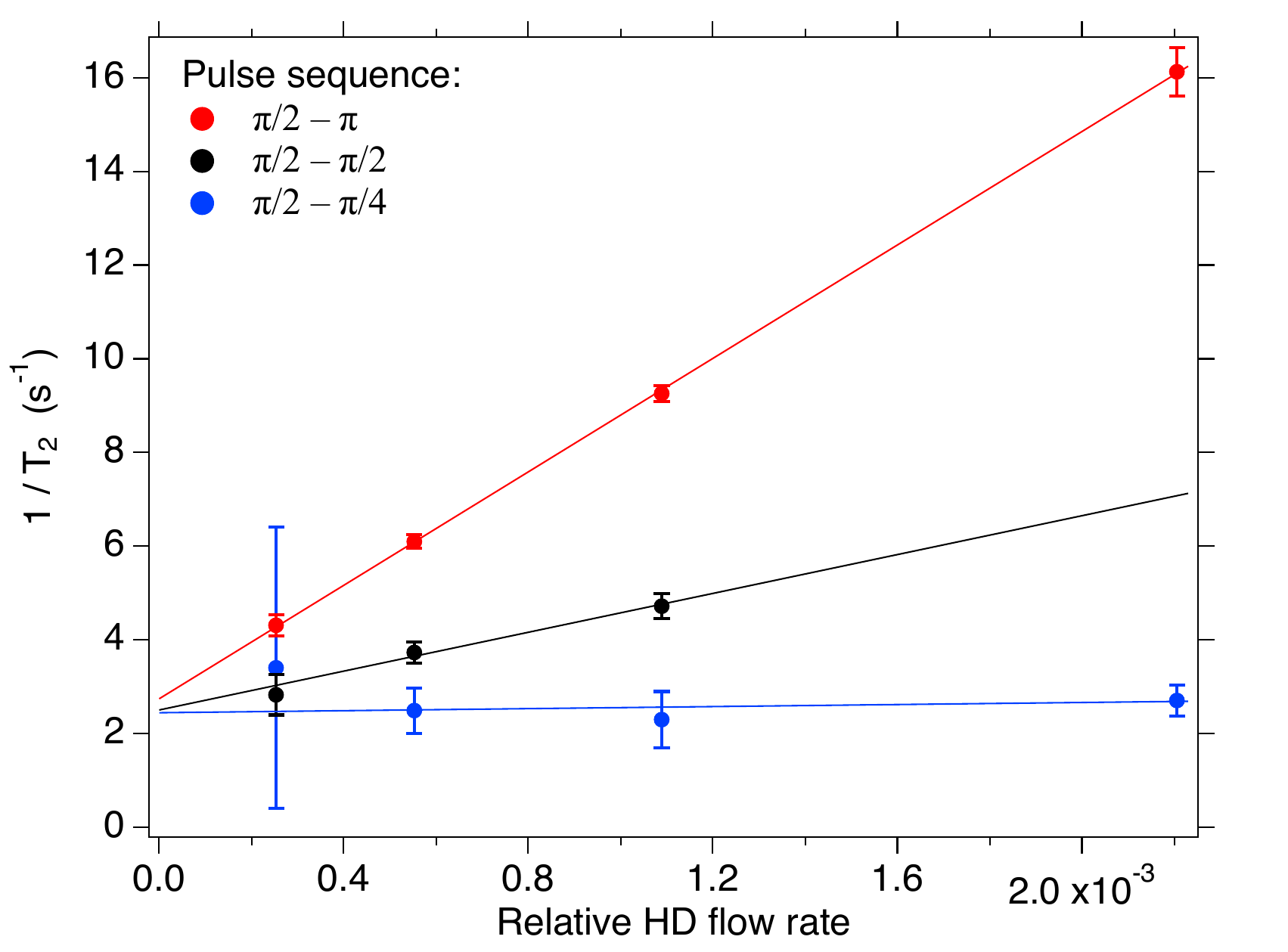}
	\caption[]{ \label{fig:T2vsHD}
	   The spin-echo $T_2$ --- plotted as $1/T_2$, the decoherence rate --- as a function of the relative HD flow rate: the ratio of the gas flow rates of HD and parahydrogen during sample growth. Data is shown for four samples, all with an orthohydrogen fraction of $1 \times 10^{-4}$, and for three different spin echo sequences. The lines are a linear fit to the data. 
       We note the HD fraction in the sample is smaller than the relative HD flow rate (based on NMR FID amplitudes).
				}
\end{figure}

$T_2$ can be limited by the interactions of HD with the parahydrogen matrix itself or with impurities in the matrix: orthohydrogen, HD molecules, or other unknown impurities. At low orthohydrogen fractions, the HD molecules play a major role in limiting $T_2$. We vary the HD density and extrapolate to a ``zero-dopant'' limit, as shown in Fig. \ref{fig:T2vsHD}.

At high HD densities, $T_2$ depends strongly on the refocusing pulse: the homonuclear magnetic interactions between HD molecules experiences no cancellation for a perfect $\pi$ refocusing pulse, but are suppressed for imperfect refocusing pulses. As seen in Fig. \ref{fig:T2vsHD}, all three pulse sequences extrapolate to consistent values for $T_2$ in the limit of zero HD.

Because we reduce the relative HD flow by increasing the parahydrogen flow --- which should lower the density of both the HD dopants and any other species inadvertently deposited during sample growth --- the extrapolated limit should be the zero-density limit for both HD and any other impurities but orthohydrogen.

The measured $T_2$ coherence times --- both measured directly and the extrapolated zero-dopant values --- are shown alongside our $T_2^*$ measurements in Fig. \ref{fig:T2sVsOrtho}.

\subsection{Coherence times}
\label{sec:coherencetimes}

\begin{figure}[!ht]			\includegraphics[width=\linewidth]{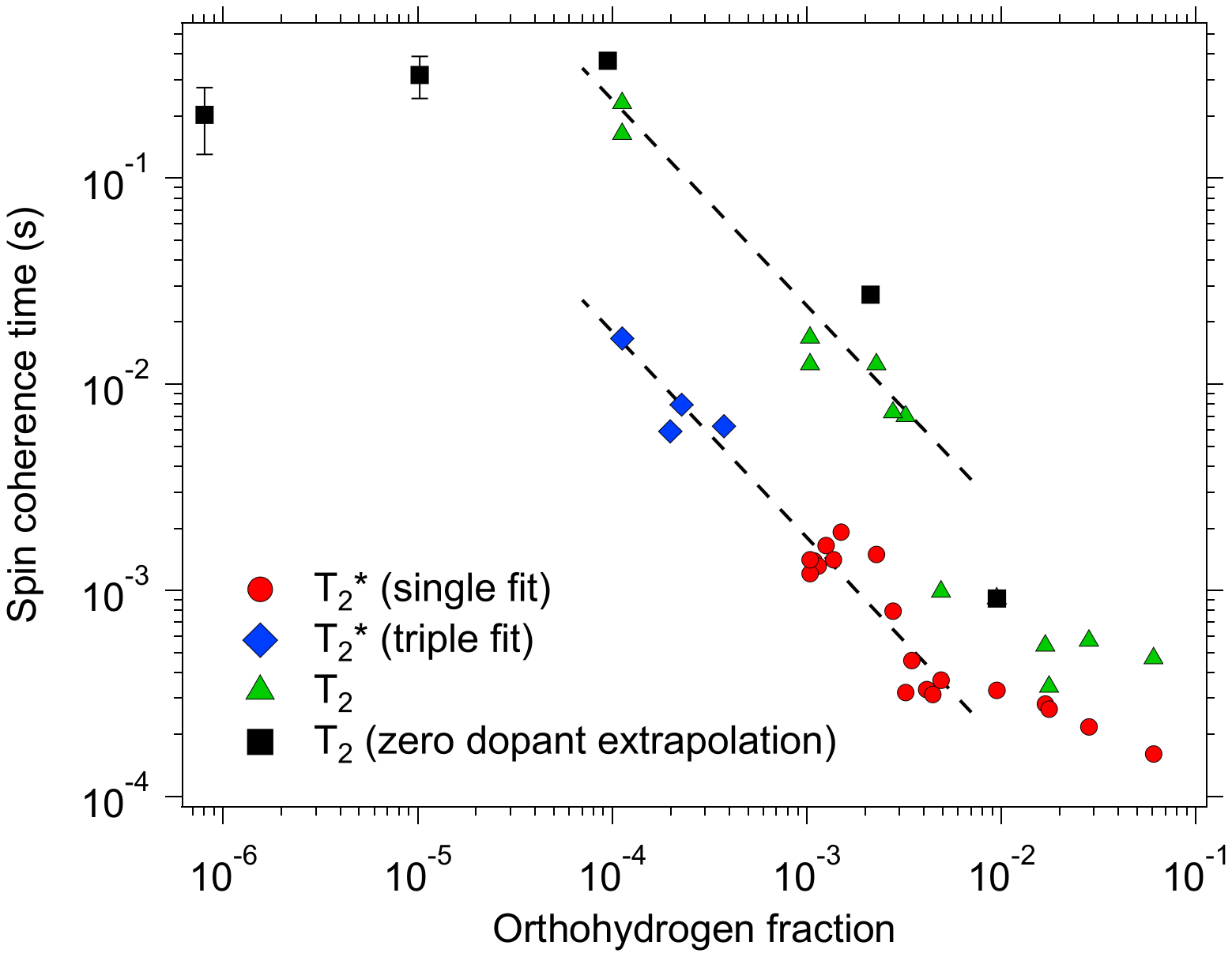}
	\caption[]{ \label{fig:T2sVsOrtho}
	$T_2^*$ and $T_2$ as a function of the orthohydrogen fraction $X$, as discussed in the text. The dashed lines are proportional to $X^{-1}$ and are included as a guide to the eye. The orthohydrogen fraction is determined from the mean temperature of the catalyst during deposition \cite{doi:10.1063/5.0049006}. The catalyst temperature varies during deposition, with typical standard deviations of $\sim 2\%$.
    }
\end{figure}

A summary of our $T_2^*$ and $T_2$ measurements are shown in Fig. \ref{fig:T2sVsOrtho}.
At high orthohydrogen fractions of $X \gtrsim 4 \times 10^{-3}$ we see complex behavior for $T_2^*$ and $T_2$.
As this complicated high-impurity regime has been explored in prior work in the field \cite{delrieu1981quantum, washburn1981nmr,  zhou1990nmr, kisvarsanyi1992nuclear}, we neglect it here for brevity. We focus on the simpler behavior observed in high-purity parahydrogen matrices of $X \lesssim 3 \times 10^{-3}$.

For $1 \times 10^{-4} \lesssim X \lesssim 3 \times 10^{-3}$, we see that as the orthohydrogen fraction $X$ decreases, the coherence times increase. 
From this behavior we conclude that ---for this range of $X$ --- the orthohydrogen impurities are the dominant limit on both $T_2$ and $T_2^*$. 
We observe that the coherence times scale as $1/X$, as expected for dilute magnetic impurities \cite{kittel1953dipolar}.  
We note that 
the spin-echo $T_2$  is consistently longer than $T_2^*$. 
This is not surprising, as $T_2^*$ is limited by the static inhomogenous fields produced by the magnetic orthohydrogen impurities, while $T_2$ is not.

Below $X{\sim}10^{-4}$, $T_2$ plateaus at $ \sim 0.3$~s. 
For measurements of $T_2^*$, the extremely long $T_1$ times at these orthohydrogen fractions make shimming the magnetic field impractical, so we cannot accurately measure $T_2^*$. However, the measured $T_2$ times provide an upper limit on $T_2^*$, so we infer that $T_2^*$  plateaus at some unknown value between $0.02$ and $0.3$~s. The mechanism by which the matrix limits $T_2$ and $T_2^*$ in the low-orthohydrogen limit is not yet understood.

\subsection{T1}

\begin{figure}[!ht]
			\includegraphics[width=\linewidth]{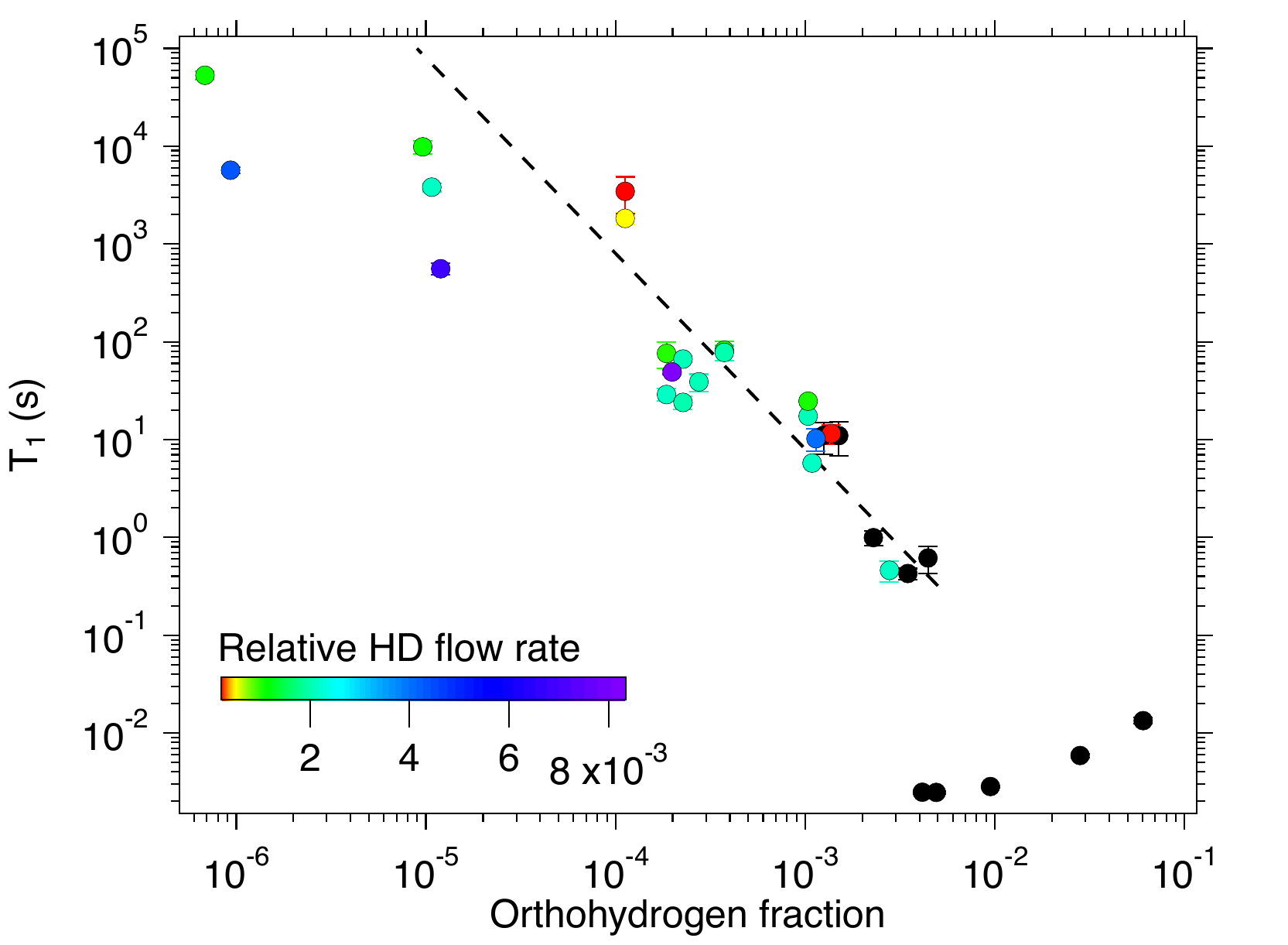}
				\caption[]{ \label{fig:T1}
				$T_1$ as a function of the orthohydrogen fraction $X$.
                At high values of $X$, the HD naturally present in the hydrogen is sufficient that additional doping with HD is not needed; undoped samples are shown as black points. For $X \lesssim 10^{-3}$, the catalyst suppresses the HD and no signal can be observed without additional doping.
                The dashed line is proportional to $X^{-2}$ and is included as a guide to the eye.
                }
\end{figure}

We measure the longitudinal relaxation time $T_1$ using a saturation–recovery sequence. First, a train of long  pulses drives the longitudinal magnetization to zero. After a delay of  $t$, we perform a standard FID measurement. We fit the amplitude of the FID power spectrum to $A - B \exp(-t/T_1)$ to extract $T_1$ \cite{T1footnote}.  
The results are shown in Fig. \ref{fig:T1}.

As with the coherence times, we concentrate on the simple behavior seen at low orthohydrogen fractions of $X \lesssim 3 \times 10^{-3}$; the anomolous behavior at high  $X$  has been previously reported  \cite{PhysRevB.46.695}. Over the range of $1 \times 10^{-4} \lesssim X \lesssim 3 \times 10^{-3}$, we see that $T_1$ becomes longer as the orthohydrogen fraction decreases.  The scaling differs from the coherence times, with $T_1$ scaling roughly as $X^{-2}$. 
We attribute this scaling to Fermi's golden rule. 
The magnitude of the magnetic field $B$ from  orthohydrogen impurities scales as $X$ \cite{kittel1953dipolar}. From Fermi's golden rule, the transition rate from a $\mu \cdot B$ interaction would be expected to scale as $X^2$. 

For $X\lesssim 10^{-4}$, $T_1$ continues to increase with decreasing $X$, but no longer follows the $X^{-2}$ scaling. However, we observe that $T_1$ depends strongly on the HD density, indicating that $T_1$ is still primarily limited by impurities in the matrix.

\section{Conclusions and future work}

The measured nuclear spin coherence times are promising.
Both $T_2$ and $T_2^*$ are longer than  typical values for molecular beams (which are limited by transit-time broadening),  with much larger numbers of molecules than typical beams \cite{doi:10.1021/cr200362u}.
$T_2$  and $T_2^*$ are significantly shorter than can be achieved with molecules in the liquid phase \cite{slichter1996principles} or in optical traps \cite{doi:10.1126/science.aal5066, lin2022seconds}, but parahydrogen offers significantly colder temperatures than the former and significantly larger numbers of molecules than the latter.
The longest values of $T_2^*$ observed for HD in parahydrogen are comparable to state-of-the-art values (for protons) in solid state NMR experiments that employ magic-angle-spinning and  nuclear-decoupling pulse sequences
\cite{REIF20121, SIMOESDEALMEIDA2023107557, doi:10.1126/sciadv.adx6016}.

Unfortunately, the extremely long $T_1$ times at low orthohydrogen fractions make it impractical to take data using traditional NMR techniques. Fortunately, techniques exist to restore nuclear spin polarization on faster timescales than $T_1$. One promising approach is to implant a second species of molecule with a metastable paramagnetic state that can be optically excited. Prior work has shown that dynamical nuclear polarization techniques can be used to transfer the spin polarization of the electron spin of metastable paramagnetic molecules to the nuclear spin of the species of interest \cite{henstra1990high, eichhorn2013high, eichhorn2013apparatus}. Once the nuclear spin polarization has been established, the optically-excited metastable states relax to diamagnetic ground states, restoring a low-magnetic-noise environment. This technique can provide nuclear spin polarizations that are significantly larger than thermal equilibrium on a timescale independent of $T_1$ \cite{henstra1990high, eichhorn2013high, eichhorn2013apparatus}.

In the dilute regime explored in this work, the limits on coherence times from orthohydrogen impurities arise from long-range magnetic dipole--dipole interactions. Thus, we expect these results to be quite general, simply scaling with the gyromagnetic ratio. In contrast, the limitations from the parahydrogen matrix itself may depend on the detailed properties of the trapped molecule, as seen before with atoms \cite{PhysRevA.100.063419}.  We hope to understand these limits by comparing different NMR-active diamagnetic molecules. By examining the dependence (or lack thereof) on  molecular properties --- such as rotational constant, gyromagnetic ratio, J-coupling, and nuclear spin-rotation coupling --- we would expect to gain an understanding of this physics.

Rotational averaging of the intramolecular magnetic dipole-dipole interaction is essential to achieve long $T_2^*$ times for molecules with more than one nuclear spin.
As known from IR spectra, HD is not the only molecule to rotate when trapped in solid parahydrogen \cite{Momose1998}. An important question to answer is what range of molecules will have this motional averaging; certainly there is some molecular size scale for which this will no longer hold. 

Finally, we hope to explore the properties of diamagnetic molecules relevant for searches for symmetry-violating new physics \cite{chen2024relativistic}.

\section*{Acknowledgements}
This material is based upon work supported by the National Science Foundation under Grant No. PHY-2309280.
We gratefully acknowledge helpful conversations with David Patterson and Amar Vutha.

\bibliography{NuclearSpinCoherence2025}

\begin{thebibliography}{49}
\bibitem [{T2f()}]{T2footnote}%
  \BibitemOpen
  \href@noop {} {}\bibinfo {note} {The factor of two is to account for the
  standard definition of $T_2$, which is the exponential timescale for the
  relaxation of the transverse magnetization. We are fitting to the FID power,
  which is proportional to the square of the transverse
  magnetization.}\BibitemShut {Stop}%
\bibitem [{T1f()}]{T1footnote}%
  \BibitemOpen
  \href@noop {} {}\bibinfo {note} {We note the $T_1$ obtained by this method is
  slightly different than the standard convention, where $T_1$ describes the
  rate of change of the longitudinal magnetization. Because the power spectrum
  is proportional to the square of the magnetization, this results in
  discrepancies from the standard convention that are $\lesssim
  40\%$.}\BibitemShut {Stop}%
\end{thebibliography}

\end{document}